\documentstyle[12pt]{article}

\def\be{\begin{equation}}
\def\ee{\end{equation}}
\def\bea{\begin{eqnarray}}
\def\eea{\end{eqnarray}}

\def\[{\lfloor{\hskip 0.35pt}\!\!\!\lceil}
\def\]{\rfloor{\hskip 0.35pt}\!\!\!\rceil}

\def\IR{\relax{\rm I\kern-.18em R}}
\def\IC{\relax{\rm I\kern-.18em C}}

\newcommand{\AmS}{{\protect\the\textfont2
  A\kern-.1667em\lower.5ex\hbox{M}\kern-.125emS}}

\catcode`@=11
\def\un#1{\relax\ifmmode\@@underline#1\else
        $\@@underline{\hbox{#1}}$\relax\fi}
\catcode`@=12

\def\fracm#1#2{\hbox{\large{${\frac{{#1}}{{#2}}}$}}}

\def\ad{{\kern0.5pt
                   \alpha \kern-5.05pt
\raise5.8pt\hbox{$\textstyle.$}\kern
0.5pt}}

\def\Dot#1{{\kern0.5pt
     {#1} \kern-5.05pt \raise5.8pt\hbox{$\textstyle.$}\kern
0.5pt}}

\def\a{\alpha}
\def\b{\beta}

\def\e{\epsilon}

\def\g{\gamma}

\def\bo{{\raise.15ex\hbox{\large$\Box$}}}               
\def\pa{\partial}                                       
\def\TH{{\raise.2ex\hbox{$\displaystyle \bigodot$}\mskip-4.7mu \llap H
\;}}
\def\face{{\raise.2ex\hbox{$\displaystyle \bigodot$}\mskip-2.2mu \llap
{$\ddot
        \smile$}}}                                      

\def\Tilde#1{\widetilde{#1}}                    
\def\Hat#1{\widehat{#1}}                        
\def\Bar#1{\overline{#1}}                       
\def\leftrightarrowfill{$\mathsurround=0pt \mathord\leftarrow \mkern-6mu
        \cleaders\hbox{$\mkern-2mu \mathord- \mkern-2mu$}\hfill
        \mkern-6mu \mathord\rightarrow$}
\def\dvec#1{\vbox{\ialign{##\crcr
        \leftrightarrowfill\crcr\noalign{\kern-1pt\nointerlineskip}
        $\hfil\displaystyle{#1}\hfil$\crcr}}}           

\def\fracm#1#2{\hbox{\large{${\frac{{#1}}{{#2}}}$}}}
\def\frac#1#2{{\textstyle{#1\over\vphantom2\smash{\raise.20ex
        \hbox{$\scriptstyle{#2}$}}}}}                   
\def\sfrac#1#2{{\vphantom1\smash{\lower.5ex\hbox{\small$#1$}}\over
        \vphantom1\smash{\raise.4ex\hbox{\small$#2$}}}} 
\def\bfrac#1#2{{\vphantom1\smash{\lower.5ex\hbox{$#1$}}\over
        \vphantom1\smash{\raise.3ex\hbox{$#2$}}}}       
\def\afrac#1#2{{\vphantom1\smash{\lower.5ex\hbox{$#1$}}\over#2}}    

\newskip\humongous \humongous=0pt plus 1000pt minus 1000pt
\def\caja{\mathsurround=0pt}
\def\eqalign#1{\,\vcenter{\openup2\jot \caja
        \ialign{\strut \hfil$\displaystyle{##}$&$
        \displaystyle{{}##}$\hfil\crcr#1\crcr}}\,}
\newif\ifdtup

  \def\pp{{\mathchoice
          {
              \kern 1pt%
              \raise 1pt
              \vbox{\hrule width5pt height0.4pt depth0pt
                    \kern -2pt
                    \hbox{\kern 2.3pt
                          \vrule width0.4pt height6pt depth0pt
                          }
                    \kern -2pt
                    \hrule width5pt height0.4pt depth0pt}%
                    \kern 1pt
           }
            {
              \kern 1pt%
              \raise 1pt
              \vbox{\hrule width4.3pt height0.4pt depth0pt
                    \kern -1.8pt
                    \hbox{\kern 1.95pt
                          \vrule width0.4pt height5.4pt depth0pt
                          }
                    \kern -1.8pt
                    \hrule width4.3pt height0.4pt depth0pt}%
                    \kern 1pt
            }
            {
              \kern 0.5pt%
              \raise 1pt
              \vbox{\hrule width4.0pt height0.3pt depth0pt
                    \kern -1.9pt  
                    \hbox{\kern 1.85pt
                          \vrule width0.3pt height5.7pt depth0pt
                          }
                    \kern -1.9pt
                    \hrule width4.0pt height0.3pt depth0pt}%
                    \kern 0.5pt
            }
            {
              \kern 0.5pt%
              \raise 1pt
              \vbox{\hrule width3.6pt height0.3pt depth0pt
                    \kern -1.5pt
                    \hbox{\kern 1.65pt
                          \vrule width0.3pt height4.5pt depth0pt
                          }
                    \kern -1.5pt
                    \hrule width3.6pt height0.3pt depth0pt}%
                    \kern 0.5pt
            }
        }}

  \def\mm{{\mathchoice
                       {
                             \kern 1pt
               \raise 1pt    \vbox{\hrule width5pt height0.4pt depth0pt
                                  \kern 2pt
                                  \hrule width5pt height0.4pt depth0pt}
                             \kern 1pt}
                       {
                            \kern 1pt
               \raise 1pt \vbox{\hrule width4.3pt height0.4pt depth0pt
                                  \kern 1.8pt
                                  \hrule width4.3pt height0.4pt depth0pt}
                             \kern 1pt}
                       {
                            \kern 0.5pt
               \raise 1pt
                            \vbox{\hrule width4.0pt height0.3pt depth0pt
                                  \kern 1.9pt
                                  \hrule width4.0pt height0.3pt depth0pt}
                            \kern 1pt}
                       {
                           \kern 0.5pt
             \raise 1pt  \vbox{\hrule width3.6pt height0.3pt depth0pt
                                  \kern 1.5pt
                                  \hrule width3.6pt height0.3pt depth0pt}
                           \kern 0.5pt}
                       }}

\def\pd{{\kern0.5pt
                   + \kern-5.05pt \raise5.8pt\hbox{$\textstyle.$}\kern
0.5pt}}

\def\pmd{{\kern0.5pt
                  \pm \kern-5.05pt \raise6.3pt\hbox{$\textstyle.$}\kern1.5pt}}

\def\md{{\mathchoice
   {
      {{\kern 1pt - \kern-6.2pt \raise5pt\hbox{$\textstyle.$}\kern 1pt}}}
    {
      {{\kern 1pt - \kern-6.2pt \raise5pt\hbox{$\textstyle.$}\kern 1pt}}}
    {
      {\kern0.5pt - \kern-5.05pt \raise3.4pt\hbox{$\textstyle.$}\kern0.5pt}}
    {
      {\kern0.5pt - \kern-5.05pt \raise3.4pt\hbox{$\textstyle.$}\kern0.5pt}}}}

\def\ad{{\dot{\alpha}}}

\def\pp{{\mathchoice
          {
              \kern 1pt%
              \raise 1pt
              \vbox{\hrule width5pt height0.4pt depth0pt
                    \kern -2pt
                    \hbox{\kern 2.3pt
                          \vrule width0.4pt height6pt depth0pt
                          }
                    \kern -2pt
                    \hrule width5pt height0.4pt depth0pt}%
                    \kern 1pt
           }
            {
              \kern 1pt%
              \raise 1pt
              \vbox{\hrule width4.3pt height0.4pt depth0pt
                    \kern -1.8pt
                    \hbox{\kern 1.95pt
                          \vrule width0.4pt height5.4pt depth0pt
                          }
                    \kern -1.8pt
                    \hrule width4.3pt height0.4pt depth0pt}%
                    \kern 1pt
            }
            {
              \kern 0.5pt%
              \raise 1pt
              \vbox{\hrule width4.0pt height0.3pt depth0pt
                    \kern -1.9pt  
                    \hbox{\kern 1.85pt
                          \vrule width0.3pt height5.7pt depth0pt
                          }
                    \kern -1.9pt
                    \hrule width4.0pt height0.3pt depth0pt}%
                    \kern 0.5pt
            }
            {
              \kern 0.5pt%
              \raise 1pt
              \vbox{\hrule width3.6pt height0.3pt depth0pt
                    \kern -1.5pt
                    \hbox{\kern 1.65pt
                          \vrule width0.3pt height4.5pt depth0pt
                          }
                    \kern -1.5pt
                    \hrule width3.6pt height0.3pt depth0pt}%
                    \kern 0.5pt
            }
        }}

  \def\mm{{\mathchoice
                       {
                             \kern 1pt
               \raise 1pt    \vbox{\hrule width5pt height0.4pt depth0pt
                                  \kern 2pt
                                  \hrule width5pt height0.4pt depth0pt}
                             \kern 1pt}
                       {
                            \kern 1pt
               \raise 1pt \vbox{\hrule width4.3pt height0.4pt depth0pt
                                  \kern 1.8pt
                                  \hrule width4.3pt height0.4pt depth0pt}
                             \kern 1pt}
                       {
                            \kern 0.5pt
               \raise 1pt
                            \vbox{\hrule width4.0pt height0.3pt depth0pt
                                  \kern 1.9pt
                                  \hrule width4.0pt height0.3pt depth0pt}
                            \kern 1pt}
                       {
                           \kern 0.5pt
             \raise 1pt  \vbox{\hrule width3.6pt height0.3pt depth0pt
                                  \kern 1.5pt
                                  \hrule width3.6pt height0.3pt depth0pt}
                           \kern 0.5pt}
                       }}

\def\pd{{\kern0.5pt
                   + \kern-5.05pt \raise5.8pt\hbox{$\textstyle.$}\kern
0.5pt}}

\def\pmd{{\kern0.5pt
                  \pm \kern-5.05pt \raise6.3pt\hbox{$\textstyle.$}\kern1.5pt}}

\def\md{{\mathchoice
   {
      {{\kern 1pt - \kern-6.2pt \raise5pt\hbox{$\textstyle.$}\kern 1pt}}}
    {
      {{\kern 1pt - \kern-6.2pt \raise5pt\hbox{$\textstyle.$}\kern 1pt}}}
    {
      {\kern0.5pt - \kern-5.05pt \raise3.4pt\hbox{$\textstyle.$}\kern0.5pt}}
    {
      {\kern0.5pt - \kern-5.05pt \raise3.4pt\hbox{$\textstyle.$}\kern0.5pt}}}}

\def\dslash{\not{\hbox{\kern-2pt $\partial$}}}
\def\Dslash{\not{\hbox{\kern-4pt $D$}}}
\def\pslash{\not{\hbox{\kern-2.3pt $p$}}}
 \newtoks\slashfraction
 \slashfraction={.13}
 \def\slash#1{\setbox0\hbox{$ #1 $}
 \setbox0\hbox to \the\slashfraction\wd0{\hss \box0}/\box0 }
 
\font\ro=cmsy10                          
\def\kcr{{\hbox{\ro \char'170}}}                
\def\ktl{{\hbox{\ro \char'170}}}        
\def\ktr{{\hbox{\ro \char'170}}}        
\def\kbl{{\hbox{\ro \char'170}}}        
\def\kbr{{\hbox{\ro \char'170}}}        

\def\plpl{\raise-2pt\hbox{$\raise3pt\hbox{$_+$}\hskip-6.67pt\raise0.0pt
\hbox{$^+$}\hskip 0.01pt$}}
\def\mimi{\raise-2pt\hbox{$\raise3pt\hbox{$_-$}\hskip-6.67pt\raise0.0pt
\hbox{$^-$}\hskip 0.01pt$}} 

\def\bo{{\raise.15ex\hbox{\large$\Box$}}}               
\def\pa{\partial}                                       
\def\TH{{\raise.2ex\hbox{$\displaystyle \bigodot$}\mskip-4.7mu \llap H \;}}
\def\face{{\raise.2ex\hbox{$\displaystyle \bigodot$}\mskip-2.2mu \llap {$\ddot
        \smile$}}}                                      

\def\Tilde#1{\widetilde{#1}}                    
\def\Hat#1{\widehat{#1}}                        
\def\Bar#1{\overline{#1}}                       
\def\leftrightarrowfill{$\mathsurround=0pt \mathord\leftarrow \mkern-6mu
        \cleaders\hbox{$\mkern-2mu \mathord- \mkern-2mu$}\hfill
        \mkern-6mu \mathord\rightarrow$}
\def\dvec#1{\vbox{\ialign{##\crcr
        \leftrightarrowfill\crcr\noalign{\kern-1pt\nointerlineskip}
        $\hfil\displaystyle{#1}\hfil$\crcr}}}           

\def\fracm#1#2{\hbox{\large{${\frac{{#1}}{{#2}}}$}}}
\def\frac#1#2{{\textstyle{#1\over\vphantom2\smash{\raise.20ex
        \hbox{$\scriptstyle{#2}$}}}}}                   
\def\sfrac#1#2{{\vphantom1\smash{\lower.5ex\hbox{\small$#1$}}\over
        \vphantom1\smash{\raise.4ex\hbox{\small$#2$}}}} 
\def\bfrac#1#2{{\vphantom1\smash{\lower.5ex\hbox{$#1$}}\over
        \vphantom1\smash{\raise.3ex\hbox{$#2$}}}}       
\def\afrac#1#2{{\vphantom1\smash{\lower.5ex\hbox{$#1$}}\over#2}}    

\parskip=\medskipamount                 
\thicklines                         

\def\oldheadpic{                                
        \setlength{\unitlength}{.4mm}
        \thinlines
        \par
        \begin{picture}(349,16)
        \put(325,16){\line(1,0){4}}
        \put(330,16){\line(1,0){4}}
        \put(340,16){\line(1,0){4}}
        \put(335,0){\line(1,0){4}}
        \put(340,0){\line(1,0){4}}
        \put(345,0){\line(1,0){4}}
        \put(329,0){\line(0,1){16}}
        \put(330,0){\line(0,1){16}}
        \put(339,0){\line(0,1){16}}
        \put(340,0){\line(0,1){16}}
        \put(344,0){\line(0,1){16}}
        \put(345,0){\line(0,1){16}}
        \put(329,16){\oval(8,32)[bl]}
        \put(330,16){\oval(8,32)[br]}
        \put(339,0){\oval(8,32)[tl]}
        \put(345,0){\oval(8,32)[tr]}
        \end{picture}
        \par
        \thicklines
        \vskip.2in}
\def\oldtitle#1#2#3#4{\oldheadpic\begin{center}\vglue.5in{\large\bf #1}\\[.6in]
        {#2}\\[.1in] {\it Department of Physics and Astronomy}\\
        {\it University of Maryland, College Park, MD 20742}\\[.6in]
        Physics Publication \#{#3}\\ {#4}\\[1.5in] {\bf ABSTRACT}\\[.1in]
        \end{center} \begin{quotation}}                 
\def\oldTitle#1#2#3#4#5#6#7{\oldheadpic\begin{center} \vglue .4in
        {\large\bf #1}\\[.4in]
        {#2}\\[.1in] {\it Department of Physics and Astronomy}\\
        {\it University of Maryland, College Park, MD 20742}\\[.1in]
        {#3}\\[.1in] {\it {#4}}\\ {\it {#5}}\\[.4in]
        Physics Publication \#{#6}\\ {#7}\\[.5in] {\bf ABSTRACT}\\[.1in]
        \end{center} \begin{quotation}}                 
\def\border{                                            
        \setlength{\unitlength}{1mm}
        \newcount\xco
        \newcount\yco
        \xco=-21
        \yco=12
        \begin{picture}(140,0)
        \put(\xco,\yco){$\ktl$}
        \advance\yco by-1
        {\loop
        \put(\xco,\yco){$\kcr$}
        \advance\yco by-2
        \ifnum\yco>-240
        \repeat
        \put(\xco,\yco){$\kbl$}}
        \xco=158
        \yco=12
        \put(\xco,\yco){$\ktr$}
        \advance\yco by-1
        {\loop
        \put(\xco,\yco){$\kcr$}
        \advance\yco by-2
        \ifnum\yco>-240
        \repeat
        \put(\xco,\yco){$\kbr$}}
        \put(-20,13){\tiny University of Maryland Elementary Particle
Physics University of Maryland Elementary Particle Physics University of
Maryland Elementary Particle Physics}
        \put(-20,-241.5){\tiny University of Maryland Elementary
Particle Physics University of Maryland Elementary Particle Physics
University of Maryland Elementary Particle Physics}
        \end{picture}
        \par\vskip-8mm}
\def\bordero{                                           
        \setlength{\unitlength}{1mm}
        \newcount\xco
        \newcount\yco
        \xco=-31
        \yco=12
        \begin{picture}(140,0)
        \put(\xco,\yco){$\ktl$}
        \advance\yco by-1
        {\loop
        \put(\xco,\yco){$\kclr}
        \advance\yco by-2
        \ifnum\yco>-240
        \repeat
        \put(\xco,\yco){$\kbl$}}
        \xco=151
        \yco=12
        \put(\xco,\yco){$\ktr$}
        \advance\yco by-1
        {\loop
        \put(\xco,\yco){$\kcr$}
        \advance\yco by-2
        \ifnum\yco>-240
        \repeat
        \put(\xco,\yco){$\kbr$}}
        \put(-20,12){\ooo bacdefghidfghghdhededbihdgdfdfhhdheidhdhebaaahjhhdahba

hgdedge
   hgfdiehhgdigicba}
        \put(-20,-241.5){\ooo ababaighefdbfghgeahgdfgafagihdidihiidhiagfedhadbfd

ecdcdfa
   gdcbhaddhbgfchbgfdacfediacbabab}
        \end{picture}
        \par\vskip-8mm}
\def\headpic{                                           
        \indent
        \setlength{\unitlength}{.4mm}
        \thinlines
        \par
        \begin{picture}(29,16)
        \put(165,16){\line(1,0){4}}
        \put(170,16){\line(1,0){4}}
        \put(180,16){\line(1,0){4}}
        \put(175,0){\line(1,0){4}}
        \put(180,0){\line(1,0){4}}
        \put(185,0){\line(1,0){4}}
        \put(169,0){\line(0,1){16}}
        \put(170,0){\line(0,1){16}}
        \put(179,0){\line(0,1){16}}
        \put(180,0){\line(0,1){16}}
        \put(184,0){\line(0,1){16}}
        \put(185,0){\line(0,1){16}}
        \put(169,16){\oval(8,32)[bl]}
        \put(170,16){\oval(8,32)[br]}
        \put(179,0){\oval(8,32)[tl]}
        \put(185,0){\oval(8,32)[tr]}
        \end{picture}
        \par\vskip-6.5mm
        \thicklines}
\def\title#1#2#3#4{\border\headpic {\hbox to\hsize{#4 \hfill UMDEPP #3}}\par
        \begin{center} \vglue .5in {\large\bf #1}\\[.6in]
        {#2}\\[.1in] {\it Department of Physics and Astronomy}\\
        {\it University of Maryland, College Park, MD 20742}\\[1.5in]
        {\bf ABSTRACT}\\[.1in] \end{center} \begin{quotation}}  
\def\Title#1#2#3#4#5#6#7{\border\headpic
        {\hbox to\hsize{#7 \hfill UMDEPP #6}}\par
        \begin{center} \vglue .4in {\large\bf #1}\\[.4in]
        {#2}\\[.1in] {\it Department of Physics and Astronomy}\\
        {\it University of Maryland, College Park, MD 20742}\\[.1in]
        {#3}\\[.1in] {\it {#4}}\\ {\it {#5}}\\[.5in] {\bf ABSTRACT}\\[.1in]
        \end{center} \begin{quotation}}                 
\def\endtitle{\end{quotation}\newpage}                  

\def\qd{{\kern0.5pt q \kern-5.05pt \raise5.8pt\hbox{$\textstyle.$}\kern
0.5pt}}

\begin{document}

\def\gfrac#1#2{\frac {\scriptstyle{#1}}
        {\mbox{\raisebox{-.6ex}{$\scriptstyle{#2}$}}}}
\def\gg{{\hbox{\sc g}}}
\border\headpic {\hbox to\hsize{September 2001 \hfill
{UMDEPP 02-009}}}
\par
\setlength{\oddsidemargin}{0.3in}
\setlength{\evensidemargin}{-0.3in}
\begin{center}
\vglue .1in
{\Large\bf THE FUNDAMENTAL SUPERSYMMETRY \\
CHALLENGE REMAINS\footnote{
Supported in part by National Science Foundation Grants 
PHY-01-5-23911}  } \\[.25in]
S.\ James Gates, Jr.\ , W.\ D.\ Linch III, J.\ Phillips and L.\
Rana \footnote{gatess@wam.umd.edu}
\\[0.06in]
{\it Department of Physics\\ University of Maryland\\ 
College Park, MD 20742-4111 USA}
\\[1.5in]

{\bf ABSTRACT}\\[.01in]
\end{center}
~~~~In the following, we will review the fundamental problem that 
prevents a complete understanding of a theory of supersymmetrical 
field representations and describe its possible relation to a similar
problem facing superstring/M-theory.
\begin{quotation}
{}  
${~~}$ \\[1.3in]

${~~~}$ \newline

\endtitle

\section {INTRODUCTION}  

~~~~There are some problems in mathematics that have taken
centuries to solve.  Perhaps the best known recent example 
of this was the proof of Fermat's Last Theorem.  Through the 
extraordinary insight and persistence of Andrew Wiles we now 
possess a proof to what had been a 350-year puzzle.

Is it time to wonder whether theoretical physics is capable
of generating such problems?

In in area of supersymmetry, for 1/14 as long as the Fermat Puzzle 
lasted, there has been such a problem.  Many years ago one of the 
present authors (SJG) became fascinated with the question of ``Why 
is it that in most theories involving supersymmetry, we not able 
to describe them in a way that is independent of their dynamics?''
This is an alternate statement of the notorious ``auxiliary
field problem.''

This question still does not possess any known answer.  While this 
question has been largely overlooked, we remain convinced that it 
is a key one for any theory which claims to provide a fundamental
description of our universe. The latest realm where such a deadlock
remains is superstring/M-theory. 

Any truly covariant formulation of superstring/M-theory ought to 
permit us to understand its symmetries as readily as does general 
relativity.  Thus our less than complete understanding of the 
representation theory of supersymmetry, in our opinion, is why our 
most cherished dream of a covariant formulation of a ``final'' 
theory remains out of our collective grasps.

\section{THE OFF-SHELL SUSY PROBLEM}

~~~~The statement of the problem is rather simple.  The fact that
it has remained without a general answer for over thirty years
suggests that the answer is not.  Consider a set of fields $\{ 
\varphi_i \}$ where the index $i$ (= 1,...,$s_{max}$) counts the 
number of fields and  they may be arbitrary representations of a
D-dimensional Minkowski,  Euclidean or any other signature metric
associated with a D-dimensional  space.  We introduce a variation
operation denoted by $\delta_Q(\epsilon)$  that depends on a Grassmann
parameter $\epsilon$.  The Grassmann parameter $\epsilon^{\alpha}$ 
should transform as the spinorial representation of  the D-dimensional
space.  

We say that a set of fields forms a {\em {off}}-{\em {shell}} 
representation of supersymmetry when 
\be
 \eqalign{
[ \, \delta_Q(\epsilon_1)~,~ \delta_Q(\epsilon_2) \, ] \, \varphi_i 
&=~ i \, < \epsilon_1 \g^{\un a} \epsilon_2 > \,\pa_{\un a} \, 
\varphi_i ~~~, } \label{eq:2Chp1} 
\ee
where we assume some appropriate inner product exists in the space of 
spinors in order to make a contraction meaningful on the RHS above.  
As simple as  this statement may be, it is satisfied in a very small
number of the known constructions involving supersymmetry.

Note that the definition above is independent of the issue of an action.  
As a second step, we may consider that set of fields $\varphi_i$ appear 
in a Lagrangian ${\cal L}(\varphi)$ such that under the action of the
variation $\delta_Q (\epsilon)$, the Lagrangian is changed by a total
derivative.  

It is more often the case that one starts with a set of fields $\{{\Hat 
\varphi}{}_i \}$ (where the index $i$ = 1,...,$s_{min}$) that satisfy,
\be \eqalign{
[ \, \delta_Q(\epsilon_1)~,~ \delta_Q(\epsilon_2) \, ] \, {\Hat 
\varphi}{}_i &=~ i \, < \epsilon_1 \g^{\un a} \epsilon_2 > \,
\pa_{\un a} \, {\Hat \varphi}{}_i  ~+~ \epsilon_1^{\a} \,
\epsilon_2^{\b} {\cal F}_{\a \b \, i}\,({\Hat \varphi}) ~~~,
} \label{eq:2Chp2} \ee
for some set of functions ${\cal F}_{\a \b \, i} $.  Typically, these 
functions are such that they also arise uniformly from the variation 
of some Lagrangian ${\cal L}({\Hat \varphi})$.  In this case the
representation ${\Hat \varphi} {}_i$ is said to be an ``on-shell''
representation of supersymmetry. 

Not all of the fields in the set $\{ \varphi_i \}$ propagate Cauchy 
data.  The fields which propagate Cauchy data are called ``the propagating
fields.''  In fact only a subset denoted by $\{ {\Bar \varphi}{}_i \}$ 
will do so.  If the set  $\{ {\Bar \varphi}{}_i \}$ is isomorphic to 
the set $\{{\Hat \varphi}{}_i \}$, then we say that the $\{ \varphi_i \}$
set is ``an off-shell extension'' of the latter.  The set of fields in
$\{ {\varphi}{}_i \}$ with indices $i = (s_{min} + 1) ,...s_{max}$ are
called ``auxiliary fields''.  The manner in which these appear in the
Lagrangian implies that their equations of motion are {\em {solely}}
algebraic.

The off-shell supersymmetry problem actually consists of {\em {two}} {\em 
{distinct}} but inter-related problems.  The first problem may be stated 
as Problem (A.);  \newline $~$ \newline ``Without regard to the existence
of  an action {\em {and}} with the smallest number and spin of auxiliary
fields,  for a given set of propagating fields find a set of propagating
and auxiliary  fields for which (1.) is satisfied.'' \newline $~$
\newline This problem is  one of representation theory.  It has no
relation whatsoever to dynamics and  can be studied accordingly.  There
is a second problem.  Problem (B.);  
\newline $~$ \newline ``Given the existence of a set of fields satisfying (1.) 
are these sufficient to permit the existence of a Lagrangian?''\newline $~$ 
\newline

In most interesting supersymmetrical theories these questions become obscured
by the existence of additional symmetries (local and global) .  In the presence 
of these, equation (1) becomes modified to read,
\be
 \eqalign{
[ \, \delta_Q(\epsilon_1)~,~ \delta_Q(\epsilon_2) \, ] \, \varphi_i &=~
 i \, < \epsilon_1 \g^{\un a} \epsilon_2 > \,\pa_{\un a} \, \varphi_i ~+~ 
\delta_{symmetry} \, \varphi_i
 ~~~,
} \label{eq:2Chp3} 
\ee
which complicates the analysis.

These questions are not regarded as being of great importance throughout most
of the literature.  Progresses in supersymmetry, supergravity, superstrings and
M-theory have continued without a basic resolution of these problems...apparently.

\section{OFF-SHELL SPINNING PARTICLES}

~~~~Some years ago, [1,2] we began an avenue of attack on the off-shell supersymmetry
problem by asking whether it was possible to find a large {\em {class}} of 
theories where the off-shell supersymmetry problem might be resolved?  In this
way, we were driven to study, perhaps the simplest of supersymmetrical systems,
the off-shell spinning particles (we refer the reader to [1,2] for a
more complete list of references).  Our efforts were rewarded.  Using certain
Clifford algebra representations, we were able to show that there exist a
solution to Problem (A.) and perhaps also to Problem (B.).

To understand the nature of the our proposed solution, it is first necessary
to introduce a class of real Clifford algebras.  We may denote these real $N$ 
linearly independent $d \times d$ matrices by $\g^{I}$ with $I = 1, \dots , N$ 
which satisfy
\be
{~~~~~~~} \g^{\rm I} \, \g^{\rm J} ~+~ \g^{\rm J} \, \g^{\rm I} ~=~- \,2 
\delta^{\rm I \rm J} \, {\bf I}
~~~. \label{eq:3Chp1} 
\ee
However, we are not just interested in all finding representations that
satisfy this condition.  We wish to restrict ourselves to the subset of these
algebras that also admit the existence of another matrix denoted by $Q$ that 
satisfies the relations
\be
{~~~~~Q^2~=~1~~~,~~~} \g^{\rm I} \, Q ~+~ Q \, \g^{\rm I} ~=~ 0 ~~~~.
\label{eq:3Chp2} \ee
We call this subset of the real Clifford algebras the ``${\cal {GR}}(N,d)$
algebras.''  For a fixed value of $N$, there exists a smallest value of
$d$ (denoted by $d_N$) such that one can construct the $N+1$ linearly 
matrices $\g^{I}$ and $Q$.  The relation between $d_N$ and $N$ is simply 
expressed in terms of the Radon-Hurwitz function [3]
\be \eqalign{ {~~~~~~~~~~~}
d_N ~=~ 2^{4m + 1} \, F_{\cal {RH}}(N)  ~~~,~  }
\label{eq:3Chp3} 
\ee  
that was written in tabular form in [1,2] (see these works for notational
details).  Due the result in (\ref{eq:3Chp2}) it follows that projection 
operators can be constructed
\be {~~~~~~~~~~~~~}
P_{\pm}~=~{\fracm 12}(I~\pm~Q) ~~~,~~~ \label{eq:3Chp4} 
\ee
from which it further follows that
\be {~~~~~~~~~}
P_{+} \, \g^{\rm I} \, P_{+} ~=~ 
P_{-} \, \g^{\rm I} \, P_{-} ~=~ 0 ~~~.  \label{eq:3Chp5} 
\ee
The remaining parts of the $\g^{\rm I}$ matrices may be denoted by the symbols
\be  {~~~~~}
L_I~=~P_{+}\g^{\rm I}P_{-} ~~~,~~~
R_I~=~P_{-}\g^{\rm J}P_{+}  ~~~.  \label{eq:3Chp6} 
\ee
In particular, there are two types of ``spinor'' indices associated with 
the quantities $L^{I}$ and $ R^{I}$,
\be {~~~~~~}
L^{\rm I} ~\equiv~  (L^{\rm I} )_{i \, \hat k} ~~~,~~~  R^{\rm I} 
~\equiv~ (R^{\rm I} )_{\hat k \, i} ~~~, 
\label{eq:3Chp7} 
\ee
with each type of index taking on values from 1 to $d$. 

The object $Q$ effectively plays the role of a chirality matrix.  Accordingly,
the analog of the usual ``dotted-undotted'' notation of the usual Van der
Waerden formalism may be applied to these systems. 
\be \eqalign{ {~~~~~~~}
(L^{(\rm I} \, R^{\rm J)})_{i \, k} ~=~-2\delta^{\rm I \rm J} \, \delta_{i 
\, k} ~~~, \cr
(R^{( \rm I} \, L^{\rm J)})_{\hat i \, \hat k} ~=~-2\delta^{\rm I
\rm J} \, \delta_{\hat i \, \hat k} ~~~.  }  \label{eq:3Chp8} 
\ee
The product space of all possible $\wedge$-products of the $\g^I$-matrices
is decomposed under the action of the projection operators $P_{\pm}$ into 
four sub-spaces,
\be \eqalign{
\{ {\cal U} \}~&=~\{P_{+}, \,  P_{+} \gamma^{\rm I \rm J} P_{+}~,...,
~P_{+} \gamma^{[N]} P_{+} \}\cr
\{ {\cal M} \}~&=~\{ P_{+}\gamma^{\rm I}P_{-}~,...,~P_{+}\gamma^{[N-1]}
P_{-} \}\cr
\{ {\Hat {\cal U}} \}~&=~\{ P_{-}, \,  P_{-}\gamma^{\rm I \rm J}P_{-}~
,..., ~P_{-} \gamma^{[N]}P_{-} \}\cr
\{ {\Hat {\cal M}} \}~&=~\{ P_{-}\gamma^{\rm I}P_{+}~,...,~P_{-}\gamma^{
[N-1]} P_{+} \} } \label{eq:3Chp9} 
\ee
The relevance of this discussion to the problem of the off-shell representation
of spinning particles has been suggested by the following observation in the
works of [1,2].  (There are some subtleties in the structures of ${\cal U}$,
${\cal M}$, ${\Hat {\cal U}}$ and ${\Hat {\cal M}}$ for arbitrary $N$.  These
will be discussed more completely elsewhere.)

Associate with each of the Clifford algebra elements of $ {\cal U}$ and
$\Hat {\cal M}$ a set of 1D fields.
\be
\eqalign{
{\rm F}:\{ {\cal U} \}~&\to~\{X(\tau), \, F^{\rm I \rm J}(\tau)\, ,..., \, 
F^{[N]} (\tau) ~\}\cr
{\rm F}:\{ {\Hat {\cal M}} \}~&\to~\{ \Psi^{\rm I}(\tau), \, \Lambda^{{\rm 
I}_i \, {\rm I}_2 \, {\rm I}_3}(\tau) \, ,..., \, \Lambda^{[N-1]}(\tau)~ \} ~~.
} \label{eq:3Chp10} 
\ee
In particular, the quantities $X(\tau)$ and $\Psi^I (\tau)$ may be identified
with the position vector and NSR fermions of a 1D-spinning particle model.  
The remaining fields are auxiliary fields.  We call the multiplet of fields 
in this construction, ``the universal spinning particle multiplet'' or USPM.  
We will also later introduce its canonically conjugate momentum multiplet.

Our first proposition is that the USPM and its canonically conjugate momentum
multiplet provide a representation of the algebra in (\ref{eq:2Chp1}).  To prove
this, we write a set of supersymmetry variations
\be \eqalign{
{\delta}_{Q} \, {\rm X} \, &=~ i \epsilon^{{\,}_{\rm I}} \, \Psi_{
{\,}_{\rm I}}  ~~~, \cr 
{\delta}_{Q} \, \Psi_{{\,}_{\rm I}} \, &=~ -2\, [~ \epsilon_{{\,}_{\rm 
I}} \, (\pa_{\tau} {\rm X}  ) ~+~ d^{-1} \epsilon^{{\,}_{\rm J}} (f_{
{\,}_{{\rm I}\, {\rm J}}})_i {}^j {\cal F}_j  {}^i  ~]  ~~~, \cr
{\delta}_{Q} \,  {\cal F}_i {}^{\, j}  \, &=~ i  \epsilon^{{\,}_{\rm 
I}} \, (f_{{\,}_{{\rm I}\, {\rm K}}})_i {}^j  (\pa_{\tau} \Psi_{
{\,}_{\rm K}}  ) ~+~ i \epsilon^{{\,}_{\rm K}} \, ({L}_{{\,}_{\rm 
K}})_i {}^{\hat k} \Lambda_{\hat k} {}^j    ~~~ , \cr  
{\delta}_{Q} \, \Lambda_{\hat k} {}^j  \, &=~ 2 \epsilon^{{\,}_{\rm 
K}} \,\pa_{\tau} \, [~ ({R}_{{\,}_{\rm K}})_{\hat k} {}^l {\cal F}_l 
{}^{\, j} ~+~ d^{-1} ({R}^{{\,}_{\rm  I}})_{\hat k}{}^j (f_{{\,
}_{{\rm I}\, {\rm K}}})_j {}^l  {\cal F}_l {}^{\, j}  ~]  ~~~,
} \label{eq:3Chp11}
\ee
and where $ f_{{\rm I} {\rm J}} \equiv P_+ \gamma_{{\rm I} {\rm J}} P_+$, 
${\cal F}_{i} {}^i = \left( {L}_{{\,}_{\rm I}} \right)_j {}^{\hat k} 
\Lambda_{\hat k} {}^j = 0$.  Also in writing these transformation laws, we 
have introduced Duffin-Kummer-Petiau fields denoted by ${\cal F}_{i} {}^k$ 
and $ \Lambda_{\hat k} {}^j$.  These collectively include all the auxiliary 
fields.  It now becomes an exercise to show that these supersymmetry variation 
satisfy (\ref{eq:2Chp1}) while placing no restrictions on any of the fields 
of the type present in (\ref{eq:2Chp2}).   We thus regard this as part of a 
proof that there is a solution to Problem (A.) for the spinning particle 
models when we pick $d = d_N$.

However, the solution described above does not necessarily solve problem (B.).
In fact, our studies indicate that given the USPM {\em {alone}} and in the
{\em {general}} {\em {case}}, it is {\em {not}} possible to write an action 
that leads to the appropriate equations of motion except for the cases of 
$N$ = 1,2 and 4.  In particular, it can be shown that the representation 
described by the USPM in the special cases of $N$ = 1,2 and 4 provides a 
{\em {reducible}} supersymmetry presentation.

For the values $N$ = 1,2 and 4, there is a truncation that may be performed
to obtain a smaller representation.  The reason for the exceptional
nature of these cases can be traced back to the supersymmetry variation
of the fermionic DKP field in the USPM in the special case of $N$ = 4
(the other two are its truncations).  If we set this field to zero, then
consistency of the supersymmetry implies
\be \eqalign{   {~~~~}
0 ~=~ &2 \e^{{\,}_{\rm K}} \,\pa_{\tau} \, [~ ({R}_{{\,}_{\rm K}}
)_{\hat k} {}^l {\cal  F}_l {}^{\, j} ~+~ d^{-1} ({R}^{{\,}_{\rm 
I}})_{\hat k}{}^k (f_{{\,}_{{\rm  I}\, {\rm K}}})_k {}^l  {\cal F}_l 
{}^{\, j}  ~]  ~~~.
  }  \label{eq:3Chp12}\ee 
It is a remarkable fact that there are non-vanishing solutions for ${\cal
F}_i{}^j$ in the cases only for $N$ = 1,2 and 4. For the exceptional $N $ 
= 4 case, the action for the spinning particle is of the form
\be \eqalign{   
{\cal S}^{N=4}_{ex} &=~ \int d \tau ~ \Big[ ~ 
\fracm 12 (\, {\pa}_{ \tau} {\rm X} \,) (\, {\pa}_{\tau}{\rm X} \,) 
~+~ 
i \fracm 12 \Psi_{{\,}_{\rm I}} {\pa}_{\tau} \Psi_{{\,}_{\rm I}}~+~ 
\fracm 14 F_{{\,}_{{\rm I} {\rm J}}} \, F_{{\,}_{{\rm I} {\rm  J}}} 
~\Big] ~~~, 
  }  \label{eq:3Chp13}\ee 
where the auxiliary field satisfies ${F}_{{\,}_{{\rm I} {\rm J}}} = 
\fracm 12 \xi \epsilon_{{\,}_{{\rm I} \, {\rm J} \, {\rm K} \, {L}}} 
{F}_{{\,}_{{\rm K} {L}}} $ for $\xi = \pm 1$.  The component field 
${F}_{{\,}_{{\rm I} {\rm J}}}$ is the part of ${\cal F}_i {}^j$ that 
lies in the null-space defined by (\ref{eq:3Chp12}).  The $N$ = 4  
action is invariant under the supersymmetry variations
\be \eqalign{   
{\delta}_{Q} \, {\rm X}{}^{} \, &=~ i \, \epsilon_{{\,}_{\rm I}} \, 
\Psi_{{\,}_{\rm I}}  ~~~~, \cr
~~~~ {\delta}_{Q} \, \Psi_{{\,}_{\rm I}} {}^{} &=~  - \epsilon_{{\,
}_{\rm I}} \, (\,{\pa }_{\tau}{\rm X}{}^{} \, ) ~+~ \epsilon_{{\,}_{
\rm K}} {F}_{{\,}_{ {\rm K} {\rm I}}}   ~~~~, \cr
{\delta}_{Q} \, {F}_{{\,}_{{\rm I} {\rm J}}} {}^{} \, &=~ - \, i \, 
\fracm 12 \, [~ \epsilon_{ {\,}_{\rm I}} {\pa}_{\tau} \Psi_{{\,}_{ 
{\rm J}} }{}^{} ~-~ \epsilon_{{\,}_{\rm J}} {\pa}_{\tau} \Psi_{{\,}_{ 
{\rm I}} }{}^{} ~+~ \xi \epsilon_{{\,}_{\rm K}} \epsilon_{{\,}_{{\rm I} 
{\rm J} {\rm K}{\rm L}}} (\, {\pa}_{\tau} \Psi_{{\,}_{{\rm L}}} {}^{}
\,) ~] ~~~~~. 
  }  \label{eq:3Chp14}\ee 
The $N$ = 2 exceptional truncation of this is given by
\be \eqalign{   
{\cal S}^{N=2}_{ex} &=~ \int d \tau ~ \Big[ ~ \fracm 12 (\, {\pa}_{ 
\tau} {\rm X} \,) (\, {\pa}_{\tau}{\rm X} \,) ~+~ i \fracm 12 \Psi_{
{\,}_{\rm I}} {\pa}_{\tau} \Psi_{{\,}_{\rm I}
} ~+~ \fracm 12 F \, F ~\Big] ~~~, 
  }    \label{eq:3Chp15}\ee 
with transformation laws given by
\be \eqalign{   
{\delta}_{Q} \, {\rm X}{}^{} \, &=~ i \, \epsilon_{{\,}_{\rm I}} \, 
\Psi_{{\,}_{\rm I}} {}^{} ~~~,~~~ {\delta}_{Q} \, \Psi_{{\,
}_{\rm I}} {}^{} ~=~  - \epsilon_{{\,}_{\rm I}} \, (\, {\pa}_{\tau}
{\rm X}  \, ) ~+~ \epsilon_{{\,}_{\rm I}} \, F ~~~~, \cr
{\delta}_{Q} \, F \, &=~ - i \, \epsilon_{{\,}_{  {\rm I}}} \, (\, 
{\pa}_{\tau} \Psi_{ {\,}_{\rm I}} \,)  ~~~~.
}  \label{eq:3Chp16}\ee 
Finally there is the $N$ = 1 theory
\be \eqalign{   {~~~~~~~}
{\cal S}^{N=1}_{ex} &=~ \int d \tau ~ \Big[ ~ \fracm 12 (\, {\pa}_{\tau} 
{\rm X} \,) (\, {\pa}_{\tau}{\rm X} \,) 
~+~ i \fracm 12 \Psi  {\pa}_{\tau} 
\Psi ~\Big] ~~~, 
  }   \label{eq:3Chp17}\ee 
with transformation laws given by
\be \eqalign{   {~~~~~~~~~~~~~~}
{\delta}_{Q} \, {\rm X}{}^{} &=~ i \, \epsilon \, \Psi ~~~~, \cr
{\delta}_{Q} \, \Psi &=~  -\, \epsilon \, (\, {\pa}_{\tau}{\rm X} 
\, ) ~~~~.
  }  \label{eq:3Chp18}\ee 

It can be seen in all of these explicit cases, the equations of motion
imply
\be
{~~~~~~} \pa_{\tau} \pa_{\tau} X ~=~ 0 ~~~, ~~~ \pa_{\tau} \Psi ~=~ 0  
~~~,
 \label{eq:3Chp19}\ee
where we have suppressed the O($N$) index on the NRS spinor in order to 
discuss all the cases uniformly.  For the $N$ = 2,4 cases where there
are required to be auxiliary fields present to close the algebra, their 
equations of motion imply that they should vanish on-shell.  These results 
for the equations of motion in the special cases of $N$ = 2,4 suggest 
that in the general case of arbitrary $N$ in (\ref{eq:3Chp10}) and 
(\ref{eq:3Chp11}) we should impose the conditions
\be  {~~~~~~~~~~~}
{\cal  F}_l {}^{\, j} ~=~ 0 ~~~,~~~  \Lambda_{\hat k} {}^j ~=~ 0 ~~~,
\label{eq:3Chp20}\ee
to define the on-shell theories. In the case of arbitrary $N$, however, 
there are no solutions of (\ref{eq:3Chp12}) that possess equal numbers 
of fermions and bosons.  Thus the requirement of equality of bosons and 
fermions needed for an off-shell supersymmetry is not satisfied in the
general case of arbitrary $N$ by a truncation.  In the general case, it 
is not possible to write an appropriate action, either with or without 
the truncation.

The problem of finding an off-shell representation, such as in
(\ref{eq:3Chp11}), which does not permit the writing of an appropriate
action is well-known to superfield supergravity theories.  For these
it has been found that all the conformal degrees of freedom occur
with a superfield called the ``conformal prepotential.''  Using the
conformal prepotential alone does not permit the writing of an
action whose equations of motion correspond to those of the usual 
Einstein-Hilbert action.  To do this requires additional supermultiplets 
called ``compensators.''  It is thus natural to try a similar solution
here.

In this case, we have suggested another route to obtaining an action.  
This begins with the introduction of a second supermultiplet that we 
have named  the ``Universal Spinning Particle Momentum Multiplet'' with
component fields 
$(\pi_{{\,}_{\rm I}}, \, {\mu}_i {}^{\hat k}, \, {\rm P}, \, {\cal G}_{i} 
{}^{j})$ that possess the supersymmetry variations given by
\be \eqalign{
{\delta}_{Q} \, \pi_{{\,}_{\rm I}}   \, &=~   \epsilon_{{\,}_{\rm I}} 
\, {\rm P}  ~+~ {\rm d}^{-1} \epsilon_{{\,}_{\rm K}} \left( f_{{\,
}_{{\rm K}\, {\rm I}}} \right)_j {}^{i} \, {\cal G}_{i} {}^{j}  
~~~~, \cr 
{\delta}_{Q} \, {\mu}_i {}^{\hat k}  \, &=~ -  \, \epsilon_{{\,}_{{\rm 
K}}} \left({L }_{{\,}_{\rm K}} \right)_{k} {}^{\hat k}  \, {\cal G}_{i}
{}^{k} ~ + \,  {\rm d}^{-1} \epsilon_{{\,}_{\rm K}} \left( 
{L}_{{\,}_{\rm I}} \right)_i {}^{\hat k} \, \left( f_{{\,}_{{\rm I}\,
{\rm K}}} \right)_k {}^{l} \, {\cal G}_{l} {}^k   ~~~~, \cr 
{\delta}_{Q} \, {\rm P}  \, &=~ -i \, 2 \epsilon_{{\,}_{\rm I}} \, 
{\pa}_{\tau} \pi_{{\,}_{\rm I}}    ~~~~, \cr
{\delta}_{Q} \, {\cal G}_{i} {}^{j}  \, &=~ - i \, 2 \, {\pa}_{\tau} 
\, [~ \epsilon_{{\,}_{ \rm J}} \left(  f_{{\,}_{{\rm I}\,{\rm J}}}
\right)_{i} {}^{j} \, \pi_{{\,}_{\rm I}} ~+~ \epsilon_{{\,}_{\rm K}}
\left( {R}_{{\,}_{\rm  K}} \right)_{\hat k} {}^{j} \, {\mu}_i  {}^{\hat
k}  ~]  ~~~~,
 } \label{eq:3Chp21}
\ee
and where the restrictions given by $ {\cal G}_{i} {}^i = \left({R}_{{
\,}_{\rm I}} \right)_{\hat k} {}^i {\mu}_i {}^{\hat k} = 0$ must be
satisfied.  

The use of this second representation allows us to write an action that
seems to make progress toward the problem of writing spinning particle
theories with an arbitrary degree of extended supersymmetry.
\be \eqalign{ {~~~~~~}
{\cal L} &=~ - \, [ ~ i {\rm d}^{-1} \, {\Tilde \mu}{}_i {}^{\hat k} 
{\pa}_{\tau} \mu_i {}^{\hat k} ~+~ i \, \pi_I {\pa}_{\tau}
\pi_I ~+~ \fracm 12 {\rm P}^2 ~+~ \fracm 12 {\rm d}^{-1} 
\, ( {\cal G}_i {}^j  {\cal G}_i {}^j ) ~]  \cr
&{~~~~~}\,+~  [ ~ - i \, \Psi_I \, (\,  {\pa}_{\tau} \pi_I \, ) 
~+~ {\rm P} (\, {\pa}_{\tau} {\rm X} \, )  ~+~ {\rm d}^{-1} {\cal G}_i
{}^j {\cal F}_j {}^i  ~+~ i {\rm d}^{-1} \mu_i {}^{\hat k} \Lambda_{\hat
k} {}^i  ~ ] ~~~~.
 } \label{eq:3Chp22}
\ee
Variation of this Lagrangian with respect to all of the functions that
appear in it leads to,
\be
\begin{array}{ccc}
~\delta \mu : {~~~~} & {~~~~~}  i \, 2 \pa_{\tau} \mu ~-~ i \, 
\Lambda &=~ 0 ~~~~,\\
~\delta \pi_I : {~~~~} & ~~  i \, 2 \pa_{\tau} \pi ~+~ i 
\pa_{\tau} \Psi &=~ 0 ~~~~, \cr
~\delta {\rm P} : {~~~~} & {~~~~~~~~~~}{\rm P} ~-~ \pa_{\tau} 
{\rm X} &=~ 0 ~~~~, \cr
~\delta {\cal G} : {~~~~} & {~~~~~~~~~~} {\cal G} ~-~ 
{\cal F} &=~ 0 ~~~~, \cr
~\delta \Psi : {~~~~} & ~~ {~~~~~~~~~~} - i \pa_{\tau} \pi 
&=~ 0 ~~~~, \cr
~\delta {\cal F} : {~~~~} & {~~~~~~~~~~} {~~~~~~~~~~} {\cal G} 
 &=~ 0 ~~~~, \cr
~\delta \Lambda : {~~~~} &{~~~~~~~~~~} {~~~~~~~~~~} \mu  &=~ 0 
~~~~, \cr
~\delta  {\rm X} : {~~~~} & {~~~~~~~~~~} {~~~~~~~~~~} \pa_{\tau} 
{\rm P} &=~ 0  ~~~~. 
\end{array}  
\label{eq:3Chp23}
\ee
(Once again we have suppressed the indices on the fields for
the sake of simplicity.)

Clearly in the bosonic sector we see
\be \eqalign{
 \pa_{\tau} {\rm P} &=~ 0  ~\&~ {\rm P} ~-~ \pa_{\tau} 
{\rm X} ~=~ 0 ~\to~ \pa_{\tau} \pa_{\tau} {\rm X} ~=~ 0  ~~,
\cr
{\cal F} &=~ {\cal G} ~=~ 0 ~~~,
} \label{eq:3Chp24}
\ee
and in the fermionic sector we find
\be   \eqalign{
i \, &\pa_{\tau} ( \pi ~+~ \fracm 12 \, \Psi ) ~=~ 0  ~~,~~
 - i \pa_{\tau} \pi ~=~ 0  ~~~, \cr
&~~~~~~~~\mu ~=~ \Lambda ~=~ 0  ~~~.
} \label{eq:3Chp25}
\ee
If we combine the first two equations in (\ref{eq:3Chp25}) we arrive
at the condition $ i \pa_{\tau} \Psi = 0$.  Comparing all of the results 
in (\ref{eq:3Chp24}) and (\ref{eq:3Chp25}) to those in (\ref{eq:3Chp19}) 
and (\ref{eq:3Chp20}), we see that the action in (\ref{eq:3Chp22})
succeeds in giving the correct equations of motion...with only one
possible subtlety.

This subtlety involves the first equation of (\ref{eq:3Chp25}). This
equations shows that the two functions $\Psi$ and $\pi$ can at most
differ from each other by a zero mode.  If this difference is
negligible, then the action in (\ref{eq:3Chp22}) seems to be
a suitable candidate to describe off-shell spinning particles.
If this is not the case, then additional modifications are required.
One way to attack these is to perform an analysis based on
Dirac quantization.  This is research that is presently
underway.

\section{THE N = 8 SPINNING PARTICLE/ SUPERGRAVITY SURPRISE}

~~~~It is possible to consider the case of $N$ = 8 within the
context of the spinning particle models we described in the
last section.  For these, the appropriate $\g$-matrices
have $d_N =$ 256.  Additionally, if we consider the spaces
${\cal U}$ and ${\cal M}$ defined in (\ref{eq:3Chp12}),
they possess an interesting structure.
\be \eqalign{
\{ {\cal U} \}~&=~\{P_{+}, \,  P_{+}\gamma^{{\rm I}_1 {\rm I
}_2}P_{+}~, \,  P_{+}\gamma^{{\rm I}_1  {\rm I}_2  {\rm I}_3  
{\rm I}_4}P_{+} \}  ~~~, \cr
\{ {\cal M} \}~&=~\{ P_{+}\gamma^{\rm I}P_{-}\, , ~P_{+}\gamma^{
{\rm I}_1  {\rm I}_2  {\rm I}_3}P_{-} \}  ~~~, } \label{eq:4Chp1} 
\ee
where the 4-form on the first line above is necessarily self-dual,
i.\ e.\  $f_{{\rm I}_1  {\rm I}_2  {\rm I}_3  {\rm I}_4} = \fracm 
1{4!} \epsilon _{{\rm I}_1 {\rm I}_2  {\rm I}_3  {\rm I}_4 {\rm I}_5  
{\rm I}_6  {\rm I}_7  {\rm I}_8} f_{{\rm I}_5  {\rm I}_6  {\rm I}_7  
{\rm I}_8}$ and $f_{{\rm I}_1  {\rm I}_2  {\rm I}_3  {\rm I}_4} 
\equiv P_{+}\gamma_{{\rm I}_1  {\rm I}_2  {\rm I}_3  {\rm I}_4}
P_{+}$. 

It is simple counting argument to note that the degeneracies of
the elements in ${\cal U}$ go as 1, 28, and 35.  The obtaining
of the thirty-five is due to the duality condition.  Similarly,
the degeneracies of the elements of ${\cal M}$ go as 8 and
56.  We can go through a similar argument with regards to the
elements of $\Hat {\cal U}$ and $\Hat {\cal M}$.  The only
major difference is that the 4-form in $\Hat {\cal U}$ is
anti-self-dual. 

To the veteran supergravity researcher, the numbers  1, 28, and 35
along with 8 and 56 are striking because these are very reminiscent
of the number of fields that occur in the 4D, $N$ = 8 supergravity
multiplet.  However, these numbers also refer to fields of
different spin.  So at first it seems highly unlikely that
their appearance in our present context is related to 4D, $N$ =
8 supergravity.

In order to make this connection more closely we should also note
that the numbers 1, 28, and 35 along with 8 and 56 in a sense
correspond to half of 4D, $N$ = 8 supergravity.  Since these
are associated with our one-dimensional construct, in order to
get a complete 4D, $N$ = 8 supergravity spectrum we ultimately
expect the appearance of a perhaps a two dimensional construct
similar to a string.  Similarly, since 4D, $N$ = 8 supergravity
is the toric reduction of 11D, $N$ = 1 supergravity, it is possible
that the result of this section is the herald of some connection
to M-theory.  In the past, we have conjectured there may exist
an NSR type of formulation of M-theory.  The results in this
section lend some support of this.

We have also spent some effort investigating realizations of
1D representations of supersymmetry from another viewpoint [4,5].
It is known that super Virasoro algebras make their appearance
in superstring theory.  We have also been undertaking a study
of the super Virasoro algebras based on model-independent
geometrical realizations.  We have seen how a set of vector
fields constructed over the superspace with coordinates
$(\tau , \, \zeta^I)$ naturally possesses a representation of
the centerless super Virasoro algebra.  Furthermore, [5]
by use of the co-adjoint representation, it has been seen
that generalized geometrical realization implies that there
is a relation between the order of a $p$-form that appears
as a generator and the ``spin'' $s$ of its co-adjoint field.
This relation takes the form
\be
{~~~~~~~~~~~~~~~~~~} s ~=~ \fracm 12 \, ( \, 4 ~-~ p \,) ~~~.
 \label{eq:4Chp2} 
\ee

Let us call the elements of ${\cal U}$ and $\Hat {\cal U}$
automorphic forms.  This is an appropriate name for these 
since they may be thought of as linear maps acting to map
the spaces of the definite chirality spinors of the $\g$-matrices
back into themselves.  Similarly, we can call the elements of
${\cal M}$ and $\Hat {\cal M}$ homomorphic forms because they
act as linear maps taking spinors of one chirality and
mapping them onto the space of the opposite chirality.
All of the elements of ${\cal U}$, $\Hat {\cal U}$,
${\cal M}$ and $\Hat {\cal M}$ are indeed forms.  So 
according to the relation in (\ref{eq:4Chp2}) we can
assign a ``spin'' to each of them.  This is done in the
table below.

\begin{center}
${~~~~~~~~~~~~~~~~~~~}$ ${\cal {GR}}$($8, 8$) Homomorphic \newline 
${~~~~~~~~~~~~~~~~~~~}$ and Automorphic Forms   \newline 
${~~}$

\renewcommand\arraystretch{1.2}
\begin{tabular}{|c|c|c| }\hline
 $~~{\rm {Algebraic~ Element}}~~$ & ${\rm Spin}$  & ${\rm Degeneracy}$ 
\\ \hline  \hline
$~~{\cal U}({\bf I})~~$ &  $~~2~~$ & $~~1~~$  \\ \hline
$~~{\Hat {\cal U}}({\bf I})~~$ &   $~~2~~$ & $~~1~~$  \\ \hline
$~~{\cal M}(f_{\rm I})~~$ &  $~~3/2~~$ & $~~8~~$  \\ \hline
$~~{\Hat {\cal M}}({\Hat f}_{\rm I })~~$ &   $~~3/2~~$ & $~~8~~$  \\ \hline
$~~{\cal U}(f_{{\rm I} {\rm J}})~~$ &  $~~1~~$ & $~~28~~$  \\ \hline
$~~{\Hat {\cal U}}({\Hat f}_{{\rm I} {\rm J}})~~$ &   $~~1~~$ & $~~28~~$  \\ \hline
$~~{\cal M}(f_{{\rm I}_1 {\rm I}_2 {\rm I}_3})~~$ &  $~~1/2~~$ & $~~56~~$  
\\ \hline
$~~{\Hat {\cal M}}({\Hat f}_{{\rm I}_1 {\rm I}_2 {\rm I}_3 })~~$ &   
$~~1/2~~$ & $~~56~~$  
\\ \hline
$~~{\cal U}(f^{-}_{{\rm I}_1 {\rm I}_2 {\rm I}_3 {\rm I}_4})~~$ &  
$~~0~~$ & $~~35~~$  \\ \hline
$~~{\Hat {\cal U}}({\Hat f}^{+}_{{\rm I}_1 {\rm I}_2 {\rm I}_3 {\rm I}_4
})~~$ &  $~~0~~$ & 
$~~35~~$  \\ \hline
\end{tabular}
\end{center}

Thus we seem to find the following interesting result.  Each
element in ${\cal U}$, $\Hat {\cal U}$, ${\cal M}$ and $\Hat 
{\cal M}$ can be associated with one of the fields that appears 
in 4D, $N$ = 8 supergravity!

\section{CONCLUSIONS}

~~~~We believe that this simplest of supersymmetrical systems has 
important lessons for as yet unsolved problems in the topic.  In 
closing we wish to reiterate what has been demonstrated.  Firstly, 
for spinning particle  actions, up to the issue of a zero mode, an 
off-shell formulation seems at hand.  Obtaining the off-shell 
representation of the spinning particle fields was possible because 
these fields are in one-to-one correspondence with the elements 
of certain real Clifford algebras, the ${\cal {GR}}(N,\, d)$
algebras.  

Due to this relation to Clifford algebras, we have not just found
one off-shell representation of the spinning particle model.
Instead we have found an infinite number of such representations.
The key point is that once one has constructed the minimal
off-shell representation of the $d_N \times d_N$ matrices in 
(\ref{eq:3Chp1}), there are an infinite number of increasingly
larger sets of $d \times d$ matrices that also satisfy the 
relations in (\ref{eq:3Chp2}).  These representation must be 
non-minimal representations of the spinning particles possessing 
larger and larger set of auxiliary fields.

We also find it intriguing that these real Clifford algebras make
their appearance in this way.  Although we do not have more direct
evidence to support the following conjecture, we find this
very suggestive that perhaps KO-theory, which is also based on
real Clifford algebras, is playing some role in determining the 
representation theory of spinning particles.  Additionally, from 
the discussion in the final chapter, there may be some way in which 
off-shell spinning particles,  KO-theory together with the 
representation theory of super Virasoro algebras is connected with 
the theory of 4D, $N$ = 8 supergravity.  Due to this, we are
hopeful about thepossibility of an NSR formulation of M-theory.
Perhaps it is useful to recall the 1D nature of the M(atrix)
model formulation of M-theory in this context.  But this
work is based on a non-linear realization of a Green-Schwarz
approach by way of comparsion.
Finally, we end with one more observation and conjecture.

Let us imagine that all supersymmetrical theories possess an 
off-shell representation.  To be completely clear about this
we mean an off-shell representation like that in (\ref{eq:3Chp10}).
Such an off-shell representation may {\em {not}} by itself
permit an appropriate action.  This is exactly what we saw for
the spinning particle.  We believe this to be a very 
reasonable assumption.  Why?  As is well known in
superspace approaches to supergravity, if one only places
conventional constraints on the theory, by definition the
component fields that arise are off-shell.

Now let us take such a theory and perform toroidal compactifications
on all of the bosonic coordinates {\em {except}} the temporal
one.  Effectively this will lead to a 1D theory that must,
however, maintain all of the supersymmetry apparent in the
higher dimension.  If there is a universality of the 1D representations
that includes the spinning particle ones we have described,
then it should be governed by the real Clifford algebras
we have seen in the case of the off-shell spinning particles.
In this case, the representation theory of the higher dimensional
off-shell supergravity theories are contained in the 1D theories
we have constructed in this work and it is possible that there
is some type of encoding of all off-shell supersymmetrical
theories contained in 1D supersymmetrical theories.  

On this basis, we have looked at the issue of the possibility of
an off-shell formulation of 4D, $N$ = 8 supergravity (or alternately
the 11D, $N$ = 1 supergravity limit of low energy M-theory) and 
concluded that the smallest possible representation contains 
32,768 bosons + 32,768 fermions.  We conjecture that there exists
an off-shell 11D, $N$ = 1 supergravity theory possessing these
numbers of bosonic and fermionic degrees of freedom.  Should
this off-shell 4D, $N$ = 8 (or alternately 11D, $N$ = 1) 
supergravity multiplet exist, it should prove to be the analog
of the spinning particle multiplet in (\ref{eq:3Chp10}).
Proving this will answer part of the challenge described by our title.


\end{document}